\documentclass[preprint]{aastex}

\newcommand{\nicmos}{NICMOS~}
\newcommand{\um}{${\rm \mu m}$~}
\newcommand{\mh}{${\rm H_2~}$}

\slugcomment{Accepted by AJ}

\shorttitle{Near-IR Imaging of PPNs}
\shortauthors{Su et al.}

\begin{document}

\title{HIGH-RESOLUTION NEAR-INFRARED IMAGING AND POLARIMETRY OF FOUR
  PROTO-PLANETARY NEBULAE\altaffilmark{1}}

\author{Kate Y. L. Su\altaffilmark{2,3,4}, Bruce J. Hrivnak\altaffilmark{3},
Sun Kwok\altaffilmark{2,5}, and Raghvendra Sahai\altaffilmark{6}}

\altaffiltext{1}{This work was based on observations with the NASA/ESA
Hubble Space Telescope, obtained at the Space Telescope Institute,
which is operated by AURA, Inc., under NASA contract NAS5-26555.}
\altaffiltext{2}{Department of Physics and Astronomy, University of Calgary,
Calgary, Canada T2N 1N4; kwok@iras.ucalgary.ca}
\altaffiltext{3}{Department of Physics and Astronomy, Valparaiso University,
Valparaiso, IN 46383; bruce.hrivnak@valpo.edu}
\altaffiltext{4}{Present address: Steward Observatory, University of Arizona,
Tucson, AZ 85721; ksu@as.arizona.edu}
\altaffiltext{5}{Canada Council Killam Fellow.}
\altaffiltext{6}{JPL/Caltech, 4800 Oak Grove Drive, Pasadena, CA 91109; sahai@bb8.jpl.nasa.gov}

\begin{abstract}

High-resolution near-infrared {\it HST} NICMOS (F160W, F222M) images
and polarization (2 $\mu$m) observations were made of four bipolar
proto-planetary nebulae (PPNs): IRAS 17150$-$3224, IRAS 17441$-$2411,
IRAS 17245$-$3951, and IRAS 16594$-$4656.  The first three of these are
viewed nearly edge-on, and for the first time the central stars in
them are seen.  Color maps reveal a reddened torus between the bipolar
lobes in the edge-on cases, with bluer lobes. The polarization values
are high, with maximum values ranging from 40 to 80\%.  The polarization
patterns are basically centrosymmetric, with some deviations in the
low polarization equatorial regions. For IRAS 17150$-$3224,
circumstellar arcs are seen at 1.6 $\mu$m, along with a newly-discovered
loop in the equatorial region.  Bright caps are seen at the end of the
lobes, indicating that they are not open-ended.  A distinct
point-symmetric pattern is seen in the strengths of the polarization
vectors, especially in IRAS 17150$-$3224.  {\it HST}  NICMOS
observations provide a valuable complement to the WFPC2 visible images
in deriving the basic structure of bipolar PPNs.

\end{abstract}

\keywords{circumstellar matter --- planetary nebulae: general ---
        stars: AGB and post-AGB }

\section{INTRODUCTION}

As low- and intermediate-mass stars enter the asymptotic giant
branch (AGB) of evolution, they enshroud themselves in dusty
cocoons created by heavy mass loss in the form of stellar winds.
Such winds gradually deplete the hydrogen envelopes of the stars,
exposing their hot cores and causing the stars to increase in
temperature.  As the envelopes continue to thin as the result of
nuclear burning and mass loss, the stellar temperature eventually
increases to the extent that significant ultraviolet photons are
emitted.  The onset of photoionization leads to strong line
emission from the circumstellar material, producing a planetary
nebula (PN).

In the past 25 years, significant progress has been made in the
understanding of both the evolution of the central star and the
dynamical evolution of the nebula. It is now believed that, in most
cases, the PN central star is powered by hydrogen shell burning,
and that the dynamics and the morphology of the nebula are driven by
an interacting winds process \citep [cf.][]{kwok2000}. While the
one-dimensional structures (including shells, crowns, haloes, etc.)
can be understood by the coupled stellar and nebular evolution models
\citep{sch03}, the origin of the variety of two-dimensional structures
is still an unresolved mystery. The observed morphologies of PN range
from round to elliptical to bipolar to point symmetric, whereas their
progenitors, the AGB stars, show spherical symmetry in their envelopes
\citep[e.g.,][]{neri98,sahai93,lucas92}.  Although it is certain that the
interacting winds process plays a role in the shaping of the nebula
\citep{balick87}, the exact details on how and when such shaping
occurs remain unknown.

The study of proto-planetary nebulae (PPNs), the transitional objects between
the AGB and PN phases, is crucial to an understanding of the processes that
shape PNs.   PN morphologies must be created or enhanced during this
transitional phase.  PPNs have recently been imaged in visible light with the
{\it Hubble Space Telescope} ({\it HST}) and found to display many of the
same shapes as the PNs \citep{ueta00}. Many show bipolar reflection
nebulae, some with concentric arcs and ``searchlight beams''
\citep{sahai98a,kwok98,su98,hrivnak99,hrivnak01}. In these, there
appears to be a higher density of material in the equatorial region
forming an optically-thick dust cocoon, with low-density lobes in the polar
directions. The bipolarity observed in PPNs suggests that the shaping
of PNs is established early in the transitional phase.

In this paper we present a high spatial resolution, near-infrared
{\it HST} imaging and polarization study of four PPNs, each of which
has recently been studied with high spatial resolution visible
imaging.  Three of these display a dark lane in the visible images
that obscures the central star. In the near-infrared, one can probe
the structure of the circumstellar shell closer to the central star
than can be done in visible light.  
One of our observational goals was to detect the central stars and 
determine their precise locations.  With the polarization
observations we wanted to investigate the scattering geometry, locate
the position of the central star from the polarization vectors, learn
about the properties of the dust grains, and see if there was evidence
to support the model of a circumstellar torus in the systems. The only
PPN for which a detailed high-resolution near-infrared imaging and
polarization study has been published is AFGL 2688 (Egg Nebula),
and in this case the central star was not seen in the near-infrared
\citep{sahai98b,weintraub00}. As part of our program of
high-resolution near-infrared imaging and polarimetry of PPNs,
we have also presented preliminary results for three other PPNs
\citep{sahai00}. With an increased sample, we hope to better
understand the properties of the circumstellar material of PPNs and
the cause of the development of asymmetry in PNs.

\section{OBSERVATIONS AND DATA REDUCTION}

High-resolution near-infrared observations of these four PPNs were
obtained in 1998 with the Near-Infrared Camera and Multiobject
Spectrometer (NICMOS) on board the {\it HST} under General Observer
program ID No. 7840 (PI: Kwok). NICMOS camera 2 (NIC2) was used,
which has a very high resolution
($0\farcs076\times0\farcs075~{\rm pixel^{-1}}$) and a
19\farcs5 $\times$ 19\farcs3 field of view \citep{thompson98}.
Observations were made with a broad-band {\it H} filter (F160W),
a medium-band {\it K} filter (F222M), and three 2 \um
(1.9-2.1 \um) polarizers (POL0L, POL120L, POL240L).
The data were read out using the multiple non-destructive mode (MULTIACCUM)
to obtain a high dynamic range without saturating the detector.
An observing log, including total exposure times in the individual
bandpasses, is given in Table \ref{obslog}. Each target was imaged
three successive times in each filter at different positions on the
NIC2 array, using a predefined spiral dither pattern. The data from
the bad pixels and columns and the area blocked by the coronographic
mask were compensated by the three-step dither mode. The data were
processed with the standard \nicmos CALNICA pipeline program, which
includes bias subtraction, dark subtraction, flat-field correction,
cosmic-ray removal, and flux calibration. An additional effect that
needed to be corrected is the so-called ``pedestal effect''
due to residual charge present in certain images \citep{thompson98}.
Through use of the ``unpedestal'' program developed by
Roeland van der Marel of the Space Telescope Science Institute,
most of this effect was removed.

\placetable{obslog}

The dithered images were shifted and coadded manually using median
combining. In order to combine the dithered images accurately, we
first subgridded each data set by a subsample factor of two so that
the peak position of field stars can be measured more precisely.
The offsets between dithered images were determined by comparing
the peak positions of a number of field stars and obtaining an average
offset. The precision in this method of registering the dithered
images can be seen from the typical rms values of the offsets in the
peak positions of the field stars in an image, which are 0.040 and
0.034 pixels in the X and Y directions, respectively.

The distortion of the \nicmos dewar due to cryogen expansion affected
its optical path. This caused the effective focal length, and hence the pixel
scale, to change somewhat as a function of time during the initial active
lifetime of the instrument. The plate scales were measured every two
weeks during the active lifetime; therefore, we adopted a plate scale
for each observation using the measurement that was closest to the date of
the observation. The \nicmos detector arrays were slightly tilted relative
to the focal plane; therefore, when projected onto the sky, the pixels are
actually rectangular, not square.  The X and Y pixel scale differences need
to be corrected to carry out precision astrometry or to accurately
register NICMOS
images to images taken with other instruments such as the Wide Field Planetary
Camera 2 (WFPC2) on board {\it HST}. These scale differences must also be
accounted for when analyzing the orientation of the polarization vectors.
The images were thus rescaled spatially when necessary.

To remove (or at least reduce) extended point-spread function 
(PSF) features of the NICMOS, we have deconvolved the images using an
iterative Poisson maximum-likelihood algorithm (task ``lucy'' in
IRAF). The PSFs used in the deconvolution were generated by the
TinyTim program v.4.4 \citep{krist97}. The deconvolved images were
only used for display purposes. All scientific measurements were
done based on the original images. Specific reduction and analysis
of the polarimetric data are discussed separately in Sec.
\ref{polar}.

\section{F160W AND F222M IMAGES}

Photometric measurements were made on the F160W and F222M images of the
four PPNs using the calibration values listed in Table \ref{ph_header}.
The magnitudes determined from the FITS image headers (PHOTFNU and PHOTFLAM)
are in the {\it HST} STMAG system. Because the \nicmos filter
bandpasses do not match those of ground-based filters,
it is not straightforward to derive standard infrared magnitudes.
Since there are no direct spectrophotometric observations of Vega or other
A0V standards through the \nicmos filter system, we used the flux values for
a 0.0 magnitude star (Vega) as derived by the \nicmos Instrument Definition
Team (IDT; see Table \ref{ph_header}). Standard aperture photometry
was carried out on the four objects using elliptical apertures that
approximately matched the size of the nebulae. Apertures were chosen
that were similar in size to the nebular halos measured on the
visible WFPC2 images. The several bright field stars inside the
apertures were excluded in calculating the total integrated counts.
Table \ref{ph_result} lists the results of the photometric
measurements and includes the aperture sizes. These magnitudes on
the Vega system, m$_{Vega}$(F160W) and m$_{Vega}$(F222M), are all
close ($\le$ 0.2 mag) to the ground-based {\it H} and {\it K} measurements,
respectively, for these objects.
The brightness of each of the central stars was also measured.
We used a small aperture (2 $\times$ FWHM) and applied an aperture
correction based upon the measurement of a bright star in the field
of IRAS 17150$-$3224.
These approximate values are also listed in Table \ref{ph_result}; 
we estimate the accuracy to be $\pm$0.3 to $\pm$0.4 mag, 
the main uncertainty being the contribution of the nebular flux.  

\placetable{ph_header}

\placetable{ph_result}

The published WFPC2 F606W ({$\lambda_{eff}=$} 0.58 $\mu$m) images of
IRAS 17150$-$3224 \citep{kwok98}, IRAS 17441$-$2411 \citep{su98}, and
IRAS 17245$-$3951 \citep{hrivnak99} show a prominent bipolarity in
their nebulae, with a dark lane separating two opposite lobes in each.
Similar bipolarity is also seen in these new near-infrared images.
This indicates that there exists an asymmetry in their circumstellar dust
shells, with stellar light escaping from lower density regions to give rise
to this bipolar nebulosity. To investigate further the
presence of an asymmetry around each of these PPNs, flux-ratio  or color
maps were computed. For these we included the previous WFPC2 F606W images.
These were shifted and rotated in order to align with the F160W and
F222M images. This process was done with the IRAF/IMMATCH package,
which uses three field stars that appear in all of the images as the
reference points to perform geometrical transformations.
Since the resolutions at different wavelengths and with different
filter widths are not the same, the higher-resolution
WFPC2 F606W and NIC2 F160W images were convolved with an extended Gaussian
profile to match the resolution in the NIC2 F222M image before
computing the flux-ratio maps. These images and flux-ratio maps
are presented and discussed for each object in the following subsections.

\subsection{IRAS 17150$-$3224}

Figures 1a and 1b display the deconvolved F160W and F222M images of
IRAS 17150$-$3224 (Cotton Candy Nebula).
The object consists of two bipolar reflection lobes
that have similar sizes in all three bandpasses (F606W, F160W, and F222M).
Bright ``caps'' appear at the ends of the lobes.  The presence of these sharp
caps strongly suggests that there exists a definite boundary to the material
distribution in the polar directions rather than open-ended cavities.
The bipolar lobes look like limb-brightened bubbles in the near-infrared
images. Thus the density inside the lobes must be very low; they are
essentially closed-end cavities.

\placefigure{150hk}

Superimposed on the bipolar lobes and the surrounding halo is a
series of concentric, circular arcs. Six arcs are detected in the
F160W image, and these correspond to the inner six arcs
previously detected in the WFPC2 F606W image. This correspondence
is clearly seen in the radial intensity profiles shown in
Figure \ref{150arc}. An average separation of 0\farcs51 was obtained
between the six arcs in the F160W image;  this value agrees with the
separation measured in the F606W image using a similar method.
These arcs appear to be portions of incomplete circles.

\placefigure{150arc}

The central star can clearly be seen in both near-infrared images, in
contrast to the visible image in which the lobes were separated by a dark
lane with the star obscured. The position of the central star is similar 
in both, with an average value of R.A.(J2000) = $17^h18^m19\fs78$,
decl.(J2000) = $-32\arcdeg27\arcmin21\farcs2$.
This agrees with the position of the crossing point of the searchlight
beams seen in the F606W image, R.A.(J2000) = $17^h18^m19\fs77$ and
decl.(J2000) = $-32\arcdeg27\arcmin21\farcs1$ \citep{kwok98}, and
confirms that the cavity which produces the beams is centered on
the central star.

The inner boundries of the lobes in the near-infrared and visible images
appear not to be parallel or axially-symmetric but rather open
widely to the SW side (the right side in Figure 1a and 1b).
The inner bounderies appear to be brighter than the caps or the rest of the
lobes in these near-infrared images.
A faint loop extending from the central region toward the ENE and WSW
directions is found in each of the near-infrared images; this loop
is not seen in the visible image. The loop has an approximately
elliptical shape centered at the central star, with a size of
1\farcs11$\times$0\farcs59 (semi-major axis$\times$ semi-minor axis)
and a position angle (P.A.) of 63\arcdeg. Note that the major axis of
the loop is not perpendicular to the symmetry axis of the bipolar lobes
(P.A. = 112\arcdeg). This kind of loop structure has also been seen
in some young PPNs (e.g., Hubble 5) in visible light.

Figures 1c and 1d display the color or flux-ratio maps for F160W/F606W
(logarithmic scale) and F222M/F160W (linear scale). The bipolar
reflection lobes and the faint arcs have bluer colors than the
equatorial regions and the central star. The flux ratios in
different parts of the nebula are summarized in Table \ref{fr150}.
The bluer colors in the bipolar lobes and the arcs are consistent
with the idea that these regions are viewed in light scattered off
the dust grains. The fact that the SE cap is redder than the NW cap
suggests that the SE lobe is pointed away from us; thus its light
is further reddened by passing through more of the halo. This is
consistent with the visible image in which the SE lobe is fainter
than the NW lobe.
The reflected light from the bipolar lobes is mostly singly scattered,
as deduced from the high percentage polarization seen at 2 $\mu$m (see
Section \ref{pol_dis}). This condition can be used to set constraints
on the radial extinction of starlight and the line-of-sight scattering
optical depth as a function of radius from the central star.
The scattering optical depth at any radius can be estimated by
comparing the observed scattered light intensity to the incident
stellar radiation (e.g., Sahai et al. 1999b).
The observed brightness for the brighter NW lobe at 2.22 $\mu$m is
3.5$\times 10^{-13}$ ergs cm$^{-2}$ s$^{-1}$ arcsec$^{-2}$ at the projected
radius of 2\farcs7. The incident stellar radiation at 2.22 $\mu$m
(2.85 $\times 10^{-9}$ ergs s$^{-1}$ cm$^{-2}$ $\mu$m$^{-1}$) can be
estimated using the bolometric flux of 5.71
$\times 10^{-8}$ ergs cm$^{-2}$ s$^{-1}$ \citep{kwok96} and $T_{eff}=$ 5500
K. Assuming a 1/r$^2$ density distribution, eqn (1)
from \citet{sahai99b} gives a line-of-sight optical depth of
$\tau_{los}=$2.2$\times 10^{-2}$, which is about 200 times smaller
than we observed ($\tau_{2.2}$ = 4.2) based on the flux ratio of the
whole nebula (see the calculation below). This suggests that the
integrated light from the nebula is mostly reddened due to the radial
extinction from the star to the point where it gets scattered, and
then slightly blue-ed by the scattering process.

The central star is extremely red, with approximate color indices
of ({\it V$-$H}) $>$ 9 and ({\it H$-$K}) $\sim$ 2.0.
The spectral type of the central star, based upon observations
of scattered light from the lobes, is G2~I \citep{hu93}. The intrinsic color
for an unreddened G2~I star is $(V-H)$=1.5 and $(H-K)$=0.1; thus the
central star is highly reddened. The central star shows a F222M/F160W
flux ratio of 1.2, which is significantly larger than that expected
(0.46) from its G2~I spectral type and estimated interstellar
extinction \citep[$A_V \sim$ 1.0 mag;][]{su01}. This additional
reddening can be attributed to circumstellar material, particularly
in the dense equatorial region. By comparing the expected value of
the F222M/F160W flux ratio (0.36) from an unreddened G2~I star
($T_{eff}=5500K$) to our observed value (1.2), (assuming the standard
interstellar extinction curve \citep{cardelli89} for the nebular dust),
we find an extinction optical depth at 2.2 \um of $\tau_{2.2} = 1.8$
and at 1.6 \um of $\tau_{1.6} = 3.0$ toward the central star.
However, the total observed flux from the object gives a F222M/F160W
flux ratio of 6.3, which gives
$\tau_{2.2} = 4.2$ and $\tau_{1.6} = 7.1$. The $\tau_{1.6}/\tau_{2.2}$ ratio
from the central star (1.6) is smaller than that derived  from the whole
object, suggesting that the extinction curve between 1.6 and 2.2 \um for
dust grains along the equatorial direction is flatter than the interstellar
one. This flatter wavelength dependency on the extinction curve is likely due
to larger grain sizes in the equatorial dense region than the typical value
(0.1 \um) for interstellar grains. Other PPNs, like the Egg Nebula
\citep{sahai98a} and IRAS 04296+3429 \citep{sahai99}, also show
evidence for large grains in their dense equatorial regions.

\placetable{fr150}

The flux-ratio maps reveal non-axially symmetric patterns,
with the region between the lobes appearing as a red wedge opening
to the SW (right side of Figure 1c and 1d).
This suggests that the dense equatorial region (torus) is larger in vertical
extent on this side.
The F222M/F160W flux-ratio map shows a red source at the
 location
of the central star, with additional red regions nearby on each
side. To better illustrate the structures in the central part,
three intensity cuts (0$\farcs$5 wide) through the central red
source were made near the x-axis of the F222M/F160W flux-ratio map
(Fig. \ref{150disk}). The nearby red regions appear as peaks in
these cuts, with two pairs of peaks (radii $\sim$$0\farcs5$ and
$\sim$$1\arcsec$) and perhaps a third ($\sim$$1\farcs6$) seen, in
addition to the central peak. The pair of peaks at radius
$\sim$1\arcsec~ is associated with the extended loop seen in the
near-infrared images. The red color presumably arises from the
scattering of the extremely reddened light of the central star.
The possible outermost pair of peaks located at r$\sim$1\farcs6 is
seen primarily along one cut. These peaks arise from regions where
the S/N is low in both the F160W and F222M images. It is possible
that they represent a second, outer loop, but
at present we consider the physical reality of the outermost peaks
to be uncertain. The inner pair of peaks may represent a
cross-section of the torus. However, they have different angular
separations from the central star ($\sim$0\farcs34 for NE and
$\sim$0\farcs52 for SW). If the inner pair represents the dusty
torus, the non-equal separations imply that either the star is not
at the center of the torus or the torus is not axially symmetric.
The former of these can arise if the post-AGB central star is in
orbit with a binary companion. (Such a companion would likely be
a main-sequence or white dwarf star and therefore much fainter and
not visible.) The average separation of the inner peaks translates
to 6.4$\times 10^{15}$ cm (D/1 kpc).

\placefigure{150disk}

\subsection{IRAS 17441$-$2411}

Figures 4a and 4b show the deconvolved NIC2 F160W and F222M images of
IRAS 17441$-$2411 (Silkworm Nebula). This object showed a pronounced S-shaped
nebula in the visible F606W image \citep{su98}, and this S-shaped morphology is
seen in both of the near-infrared images.
This morphology suggests point-symmetry in the nebula.
The central star, which was not seen in the visible, is clearly resolved in both
of the near-infrared images, with a position of 
R.A.(J2000) = $17^h47^m13\fs51$, decl.(J2000) = $-24\arcdeg12\arcmin51\farcs6$.
This position agrees with the center
position of the two pairs of concentric arcs seen in the F606W image,
R.A.(J2000) = $17^h47^m13\fs49$, decl.(J2000) = $-24\arcdeg12\arcmin51\farcs0$.
The nebula has a similar size in all three bands;
this is to be expected when one is seeing scattering from the dense-walled lobes.

\placefigure{441hk}

Figures 4c and 4d show the flux-ratio maps. The reflection lobes are bluer
than the equatorial region, as is the case in IRAS 17150$-$3224.
The southern lobe is redder and fainter in visible light than the northern lobe;
both of these characteristics suggests that the southern lobe is tipped away.
The detailed flux ratios in different regions are listed in Table \ref{fr441}.
The central star is very bright in the near-infrared images, resulting in
a complex PSF; hence the flux-ratio map of F222M/F160W is complicated and
full of PSF features (wings and spikes).
In an inset in Figure 4c, the intensity contours around the central star
in the F160W/F606W flux-ratio map are shown. These contours appear to
show an axisymmetric torus perpendicular to the lobes.
Thus, similar to IRAS 17150$-$3224, the color distribution in IRAS 17441$-$2411
supports the model of a circumstellar torus.
The central star appears extremely red (presumably as a result of
circumstellar extinction), with approximate values of ({\it V$-$H}) $>$ 9 and
({\it H$-$K}) $\sim$ 2.5.

\placetable{fr441}

\subsection{IRAS 17245$-$3951}

Near-infrared images of IRAS 17245$-$3951 (Walnut Nebula) are shown in
Figure \ref{245hk}.
The central star, which was not seen in the visible, is clearly resolved in both
of the near-infrared images, with a position of 
R.A.(J2000)=$17^h28^m04\fs59$, decl.(J2000)=$-39\arcdeg53\arcmin44\farcs9$.
The nebula is small and has a size in the F160W image similar to that seen in the
F606W image.  However, the nebula is not seen in the F222M image; this may be
due to the fact that the image does not go as deep as the other two and partly
perhaps to the relatively large effect of the complicated PSF of the bright central
star in the F222M image.
A jet-like structure seen in the southern lobe in the F606W image \citep{hrivnak99}
does not appear in the F160W image.

Figure \ref{245hk}c shows the F160W/F606W flux-ratio map.
The bluest colors are associated with the lobes and the reddest colors with a torus
perpendicular to the lobes.
The northern lobe is bluer than the southern lobe, again consistent with the visible
image in which the northern lobe is brighter than the southern lobe.
The central star is very red, with ({\it V$-$H}) $>$ 3.5 and
({\it H$-$K}) $\sim$ 1.6.

\placefigure{245hk}

\subsection{IRAS 16594$-$4656}

The visible F606W image of IRAS 16594$-$4656 (Water Lily Nebula;
\citet{hrivnak99}) indicates that the bipolar nebula is oriented at an 
intermediate inclination, not nearly edge-on as in the other three cases.  
The central star is seen in the F606W image, and it has been assigned 
a spectral type of B7 \citep{vandesteene00}.  It is reddened, with
({\it V$-$H}) $\sim$ 4.2 and ({\it H$-$K}) $\sim$ 0.3.  No dark lane is
visible. Near-infrared images of IRAS 16594$-$4656 are shown in 
Figure \ref{594hk}. The nebulosity in
the F160W image bears a general resemblance to the one seen in the
visible image, but does not show as much faint detail.  In the
F606W image, included for comparison (Fig. 6c), the main lobes are
along the ESE and WNW directions with some extended structures
(``petals'') along the NE and SW directions. The nebulosity seen
in F160W is also elongated along the ESE and WNW directions, and
the petals appear very faintly. Very little nebulosity is detected
in the F222M image; this is due in part to the fact that the PSF of
the bright central star masks, to some extent, the light of the nebula.
In the F160W/F606W flux-ratio map, the faint petals
stand out as having a bluer color than the rest of the nebula.

\placefigure{594hk}

\section{2 \um POLARIZATION IMAGES}
\label{polar}

\subsection{Reduction and Analysis Procedures}

The \nicmos polarization data of the four PPNs were analyzed using the algorithm
developed by \citet{hines00} to derive the Stokes images ({\it I},{\it Q},{\it U}) and the
maps of percentage polarizations ({\it p}) and position angles ($\theta$).
Before constructing the final Stokes images and percentage polarization
and position angle maps, there were several additional
steps in the data reduction.
Any sky background was removed, since this would produce a small DC offset
that would introduce some artificial polarization.
The sky values in each image were determined by examining several blank
sky areas and an average sky value was subtracted from the image.
The X and Y pixel scale differences were adjusted to produce square pixels;
otherwise, a systematic offset in polarization position angle would have been
introduced \citep{hines00}.
In order to transform the data to square pixels, the flux of each data set was
redistributed with a bicubic spline interpolation function
using a task called ``imlintran'' in IRAF.
Since the percentage polarization and position angle maps are formed by dividing
one image by another image, one must be careful to avoid a run-away error, which
can occur when dividing by a very small number with a relatively large uncertainty.
Therefore, before calculating the result maps, the three data sets were
rebinned by 2x2 pixels to smooth out small fluctuations.
Then the threshold value for a real signal was defined as 10 $\sigma_{sky}$; any signal
below 10 $\sigma_{sky}$ was considered as noise with zero percentage polarization.

The actual uncertainties in the polarization will be larger than the Poisson errors.
Additional sources in the form of systematic errors include the background
subtraction and the image alignment.   
In order to reduce possible errors in the final percentage polarization
and position angle maps, a threshold of S/N = 10 was also set
in the final Stokes images.
As the following estimations show, with this conservative threshold, the
estimated uncertainties are not important.
We first added 1 $\sigma_{sky}$ noise to the sky-subtracted images 
(POL0L, POL120L, POL240L) and found the percentage polarization to differ
only within $\pm$3\%.  We also computed a new set of polarization maps 
using un-shifted data and found that the percentage polarization differs by
less than 3\% within the the S/N=10 threshold region, although it can increase 
up to 24\% outside that boundry.
According to the on-orbit calibration of the NICMOS polarimetric capability,
a total instrumental polarization of $<$ 1\% was found \citep{hines00}.
Since these four objects are all in the general direction of the center
of the galaxy (${\it l}$: 340$\arcdeg$ to 4$\arcdeg$, ${\it b}$:
$-$3.3$\arcdeg$ to +3.0$\arcdeg$), a small amount of polarization ($\le$5\%)
may be of interstellar origin.
Therefore, we conclude that, after correcting for the projection effect,
rebinning, and setting the above thresholds,
the uncertainty in the polarization is $\sigma_{p}$ $\le 5\%$ and 
$\sigma_{\theta}$ $\le 2\arcdeg$ (the later taken from \citet{hines00}).

\subsection{Polarization Structures}

The polarization pattern for a simple reflection nebula with a central
illuminating source, such as a PPN, is expected to be centrosymmetric
about the illuminating star. Even for a slightly more complicated system
composed of a mixture of reflected central starlight and line emission
from the extended nebulosity, such as is found in a young PN, one still
expects to observe this centrosymmetric pattern.
This is because the line-emission radiation will be unpolarized and its
effect will only be to dilute the polarized scattered light
\citep{scarrott95}. The most likely near-infrared emission line in the PPN
phase is from ${\rm H_2}$,  and the NIC2 polarizer bandpass may admit as much as
10 \% of any \mh emission present at 2.12 $\mu$m. Among these four PPNs, only
IRAS 17150$-$3224 is known to have \mh emission. This dilution effect will be
discussed in \ref{pol_dis}.

Figure \ref{pol} shows the 2 \um polarization maps for the four PPNs.
In this figure, the polarization vectors are plotted as  straight lines whose lengths
represent the strength of the polarization; these are superimposed on the contour
images of the total flux of the light through the three polarizers (Stokes {\it I} image).
The distribution of the polarization vectors shows a general centrosymmetric
pattern around the central source in each of these PPNs.
This is consistent with what one would expect from a reflection nebulosity with
a central illuminating source.
Some departure from a centrosymmetry pattern is found in the equatorial regions.
This is commonly seen in the equatorial regions of bipolar reflection nebulae
\citep{kastner95,scarrott95,sahai98b}.
In these equatorial regions, the polarization is low ($<$ 10 \%), as will be discussed
below.
This suggests that multiple scattering has reduced the strength of the polarization
and provides further evidence for the presence of a dense circumstellar torus in these
PPNs.

\placefigure{pol}

In addition to the centrosymmetry, the polarization vectors also
show a strong point symmetry in the distribution of polarization
strengths around the central star. This can be seen in plots of
the polarization vectors within different ranges of percentage
polarization. IRAS 17150$-$3224 shows remarkable point symmetry,
as seen in Figure \ref{pflowa}. The lobes and the two caps are
highly polarized, with $p > 60 \%$. The maximum percentage
polarization is found at the edge of the SE lobe (the fainter lobe
in the optical image), with $p \sim 76 \%$. In general, the light
from the lobes is polarized with $p > 50 \%$ (see the combination
of Fig. 8a and 8b). Lower polarization is found along the edges of
the lobes. The areas with $p < 10 \%$ are located in the
equatorial region. For IRAS 17441$-$2411 (Figure \ref{pflowb}),
the reflection lobes typically  possess $p\sim 45 \%$
polarization. The maximum polarization is found at the bend of the
S-shape in the southern  lobe (the fainter lobe in the optical
image), with $p \sim 60 \%$. Similar to IRAS 17150$-$3224, lower
polarization is found along the edges of the lobes and the areas
associated with the lowest polarization ($p < 10 \%$) are located
in the equatorial region.

For IRAS 17245$-$3951 (Figure \ref{pflowc}), the higher polarization
($42 \% > p > 30 \%$) is found near the ends of the nebular lobes.
Again, the low polarization ($p < 10 \%$) region is
associated with the equatorial direction.
For IRAS 16594$-$4656 (Figure \ref{pflowd}), the higher polarization region
($59 \% > p > 40 \%$) is found to be associated with the edges of the nebula
and the maximum polarization is found in the eastern lobe. The vectors
also show a general centrosymmetric pattern. Since the nebula is oriented at
an intermediate inclination \citep{su01}, the torus is no longer
edge-on and the low polarization ($p < 10 \%$) region is distributed
around the central star.

\placefigure{pflowa}

\placefigure{pflowb}

\placefigure{pflowc}

\placefigure{pflowd}

\subsection{Polarization Discussion}
\label{pol_dis}

In Figure \ref{p_percent} are shown the percentage polarization
maps for the IRAS 17150$-$3224 and IRAS 17441$-$2411. One can
clearly see the highly polarized regions at the ends and the edges
of the lobes in IRAS 17150$-$3224. Note that \mh emission is
observed in IRAS 17150$-$3224 \citep{su00}. The strongest \mh
emission is seen from the polar caps, which are among the regions
with the largest polarization strength ($\sim$ 60$-$76 \%).
However, the total flux contributed by the  \mh emission is low
compared to the total flux through three polarizers ($\sim$2.3
\%). The undiluted percentage polarization should be higher by
only $\sim$1 \%.

\placefigure{p_percent}

Comparison of our {\it HST} imaging polarimetry with that obtained 
from the ground shows the advantages of the high {\it HST} spatial 
resolution when observing small nebulae such as PPNs. 
\citet{gledhill00, gledhill01} carried out ground-based 
near-infrared ({\it J}-band) imaging polarimetry for three of these 
PPNs.   They found the polarization vectors distributed in an
elliptical pattern aligned orthogonal to the polar axis of each
nebula, rather than in the general centrosymmetric pattern found
in these {\it HST} NICMOS observations. This can be quite easily
understood as the result of the lower spatial resolution (by a
factor of five) of their observations, which results in spatial
smearing of the vectors (see \citet{lucas98}, Fig. 12). 
Similarly, their values (at 1.25 $\mu$m) for the maximum observed
polarization are less than these {\it HST} values (at 2.0 $\mu$m).  
This is also likely due to spatial smearing of the polarization values 
and to the greater effect of dilution on the polarization values by the 
unpolarized light of the central stars in their lower-resolution images. 
These effects will be greatest for the small nebula of IRAS 17245$-$3951. 
(This difference in the maximum observed polarization values is
not due to the difference in wavelengths observed, since these
authors also observed some PPNs in both {\it J} and {\it K} (2.2 $\mu$m) 
and found the maximum observed polarization at {\it K} to be comparable 
to or less than that at {\it J}).

The polarization maps depend on the dust grain properties, 
in additional to the density structure of the nebulae.
We therefore use the Monte Carlo multiple-scattering models
developed for bipolar nebulae \citep{whitney93,whitney97,lucas98}
around young stellar objects (YSOs), since these are morphologically
rather similar to PPNs for which such models are not available.
With these we analyze our polarimetric images and infer rough
constraints on the dust properties of our PPNs.

First, one can get some constraints on the grain albedo.
\citet{lucas98} simulated K-band images for low albedo ($\omega_K$ = 0.22)
and high albedo ($\omega_K$ = 0.8) dust grains (see their Fig. 14).
The low albedo image shows pinched-in, bipolar-like intensity contours in
the equatorial plane for a near edge-on system, while the high albedo
image shows more elliptical-like contours.
Since IRAS 17150$-$3224 and IRAS 17441$-$2411 appear bipolar,
low albedo grains are likely to be dominant in these two.
Furthermore, a lower albedo in the K band is consistent with single scattering
of the photons, since a high albedo would lead to multiple scattering
and a lower value for the maximum observed percentage polarization.  The values
of the observed percentage polarization are very high for these PPNs.
Both the high values of the observed percentage polarization and the pinched-in
morphologies infer that low albedo grains are dominant in these PPNs. A similar
result was also found in another PPN, Roberts 22 \citep{sahai99b}.

The value of the maximum linear percentage polarization ($p_{max}$) is a function
of the grain size and thus can be used to constrain the size of the grains
present. The value of $p_{max}$ is the highest percentage level of polarization that
can be produced by single scattering, and the maximum observed value of the
percentage polarization provides a lower limit to $p_{max}$.
In Table \ref{pmax} are listed the observed maximum values of the polarization
($p_{max}$(obs)) for these four PPNs and the Egg Nebula (for comparison).
\citet{lucas98} calculated the percentage  polarization produced by
scattering though 90\arcdeg~using particles of different radii $a$
(see their Fig. 13).
They found that for spherical silicate grains (with refractive index
m = 1.7 $-$ 0.03{\it i}), $p_{max}$ is 100 \% for grains
with radii $a \lesssim$ 0.3 $\mu$m but then drops off rapidly for larger
grains.  (For a metallic dust mixture (m = 1.7 $-$ 0.6{\it i}) this size
constraint is removed since they
maintain high polarization for a wide range of sizes.)
Thus, if we assume silicate grains (silicate absorption at 9.7 $\mu$m is
observed in IRAS 17150$-$3224; \citet{kwok95}), then these high observed
maximum polarization values indicate that
the dust must be dominated by small grains in the scattering envelope
(the bipolar lobes and halo).
Table \ref{pmax} also includes the location in the bipolar nebulae where
the observed maximum polarization is found and the nebular size in visible
light. Except for IRAS 17245$-$3951, the observed maximum polarization is
found associated with the fainter lobe for each of these bipolar nebulae.
In an axisymmetric distribution of dust illuminated by a central source and
inclined to our line of sight, if the fainter lobe is more polarized than
the brighter lobe, the most likely explanation is that the grains are forward
scattering. This is because with forward scattering, the scattered flux is
weighted toward small scattered angles \citep{whitney93}.
Thus the comparison of these near-infrared images and polarization maps with
models developed for YSOs suggests that the dust in the scattering envelope
is dominated by small, forward-scattering  grains
(radii $a \lesssim$ 0.3 $\mu$m  for silicates) with low albedos.

\section{SUMMARY AND CONCLUSIONS}

A high-resolution near-infrared imaging and polarization study of four bipolar PPNs
has been carried out using NICMOS on the {\it HST}.  In the three nearly edge-on
nebulae, the central stars are seen for the first time in these near-infrared images.
(This is in contrast to the Egg Nebula, which appears similar in visible light but in which
the central star is not seen in the near-infrared.)
All four show a general centrosymmetric pattern in their polarization vectors, indicating
that one is seeing light from the central star that has been singly-scattered
from the dust grains.
The nebular size is the same in the near-infrared images as in the previous
visible {\it HST} images, consistent with scattering from dense-walled lobes.  
Color maps show that the light from the lobes is
bluer than from the central region, and the polarization images
show that the lobes are highly polarized ($>$40\% for
IRAS 17150$-$3224 and IRAS 17441$-$2411).
These all indicate that the lobes are seen almost entirely in scattered light
and that any emission sources in the lobes are faint.
The dark region between the lobes is very red in the color maps
and the polarization there is weak and deviates from the centrosymmetric pattern;
these are consistent with the presence of a dense torus that lies perpendicular to
the bipolar lobes and causes multiple scattering in the emergent light.

The limb-brightening of the lobes seen in IRAS 17150$-$3224 suggests that they
possess low density and are viewed primarily in light scattered from the cavity walls.
Furthermore, the cap structures indicate that these cavities are not
open-ended cones.  An extended loop
was discovered along the equatorial direction in this object.

In addition to the centrosymmetric polarization patterns, point-symmetric
patterns are also seen among the polarization vectors for these four PPNs.
This is particularly striking for IRAS 17150$-$3224, the largest of the edge-on
nebulae, and was seen previously in the Egg Nebula \citep{weintraub00}.  
Such point-symmetric structures resemble the BRET
(bipolar, rotating, episodic jets) phenomenon observed in PNs \citep{lopez95}, 
and suggest that the mechanism responsible for producing such
structures is already operating in PPNs.  Although it is unclear at this
time whether the BRET phenomenon is related to the multipolar lobes
observed in PN 2440 \citep{lopez98} and in low-excitation (and young)
PNs \citep{sahai98}, our present data suggest that colliminated bipolar 
outflows commence early in the post-AGB evolution.

The results in this paper increase the number of detailed high-resolution near-infrared
imaging and polarization studies carried out with the {\it HST} from one
\citep[the Egg Nebula,][]{sahai98b} to five.
Our observations confirm the basic model of PPNs in which an equatorial region of
enhanced density (torus) obscures the central star in visible light and in which light
escaping in the polar directions scatters from the walls of polar cavities carved out
by a fast wind.
These results agree with other recent observations that show that the
morphological shaping of PN occurs
 early, before the onset of photoionization.
This model can be further tested by comparing the high-resolution mid-infrared images
of PN and PPNs that are starting to be obtained with the new 8$-$10 m telescopes.

\acknowledgments
We thank Dean Hines for helpful discussions about the \nicmos polarimetric
data reduction and K.~Y.~L.~S. thanks Roeland van der Marel for assistance with the
``unpedestal'' program.
The comments from an anonymous referee have helped to improve the presentation
and are appreciated. 
B.~J.~H. and R.~S. acknowledge support from NASA through grant numbers
GO-07840.02-A and GO-07840.01-A, respectively, from the Space Telescope Science
Institute, which is operated by the Association of Universities for Research in Astronomy,
Inc., under NASA contract NAS5-26555.
R.~S. also acknowledges support from NASA through an LTSA grant (399-20-61-00-00).
S.~K. acknowledges support from the Natural Science and Engineering Research
Council of Canada and a Killam Fellowship from the Canada Council for the Arts.

\clearpage

\begin{deluxetable}{llccccc}
\tablecaption{{\it HST} NICMOS Observing Log\label{obslog}}
\tablewidth{0pt}
\tablehead{
\colhead{IRAS ID}&\colhead{Observation}& \multicolumn{5}{c}{Total Exposure Times (s)}\\ \cline{3-7}
\colhead{}&{Date}&\colhead{F160W}&\colhead{F222M}&\colhead{POL0L}&\colhead{POL120L}&\colhead{POL240L}}
\startdata
16594$-$4656  &1998 May  2&256 &288 &288 &288 &288 \\
17150$-$3224  &1998 Aug 16&448 &416 &448 &448 &448 \\
17245$-$3951  &1998 Apr 30&256 &272 &272 &272&272 \\
17441$-$2411  &1998 Mar 10&416 &416 &416 &416 &416 \\
\enddata
\end{deluxetable}


\begin{deluxetable}{ccccccc}
\tablewidth{0pt}
\tablecaption{{\it HST} NICMOS Photometric Calibration Values \tablenotemark{a}\label{ph_header} }
\tablehead{
\colhead{Filter}&\colhead{$\lambda_{eff}$}&\colhead{$\Delta \lambda$} &\colhead{PHOTFNU}&\colhead{PHOTFLAM}&\colhead{F$_{\nu}$ for 0.0 mag}&\colhead{PSF FWHM \tablenotemark{b}} \\
\colhead{}&\colhead{(\um)}&\colhead{(\um)}&\colhead{(Jy/(ADU s$^{-1}$))} & \colhead{(ergs cm$^{-2}$ s$^{-1}$ \um$^{-1}$)} & \colhead{(Jy)}&\colhead{(\arcsec)} \\
\colhead{}&\colhead{}&\colhead{}&\colhead{}& \colhead{/(ADU s$^{-1}$)}
& \colhead{}&\colhead{}
}
\startdata
F160W&  1.593 & 0.277 & 2.190E-06 & 2.589E-15 & 1083& 0.16 \\
F222M&  2.216 & 0.120 & 5.487E-06 & 3.352E-15 & 668 & 0.20 \\
\enddata
\tablenotetext{a}{The photometric calibration was done by the NICMOS IDT
(M. Rieke, private communication).}
\tablenotetext{b}{Based upon a Gaussian fit to the PSF.}
\end{deluxetable}


\begin{deluxetable}{cccccccccc}
\tablecaption{{\it HST} NICMOS Photometry Results\label{ph_result}}
\tablewidth{0pt}
\tablehead{
\colhead{IRAS ID}
&\multicolumn{6}{c}{Entire PPN}& &\multicolumn{2}{c}{Central Star}  \\
\cline{2-7} \cline{9-10} 
 &\colhead{Size\tablenotemark{a}}
 &\multicolumn{2}{c}{F160W}& &\multicolumn{2}{c}{F222M}  & 
 &\colhead{F160W}&  \colhead{F222M} \\
\cline{3-4} \cline{6-7} 
 &(\arcsec)  &m$_{ST}$\tablenotemark{b} & m$_{Vega}$\tablenotemark{c} 
 & &m$_{ST}$\tablenotemark{b} &m$_{Vega}$\tablenotemark{c}
 & & m$_{Vega}$\tablenotemark{c} & m$_{Vega}$\tablenotemark{c}
}
\startdata
16594$-$4656 & 6.15$\times$4.28 & 12.57 & 8.94  & & 13.16 & 8.28 & & 8.9 & 8.5 \\
17150$-$3224 & 8.97$\times$4.83 & 13.67 & 10.04 & & 14.13 & 9.30 & & 13.2 & 11.5 \\
17245$-$3951 & 3.07$\times$2.43 & 14.11 & 10.48 & & 14.63 & 9.76 & & 11.7 & 10.7 \\
17441$-$2411 & 5.85$\times$4.12 & 13.70 & 10.07 & & 14.20 & 9.33 & & 13.2 & 11.1 \\
\enddata
\tablenotetext{a}{Size of the elliptical aperture: major axis $\times$ minor
axis.}
\tablenotetext{b}{Calibrated magnitude on the {\it HST} STMAG system.}
\tablenotetext{c}{Standardized magnitude with respect to Vega.}
\end{deluxetable}


\begin{deluxetable}{ccccccccc}
\tablewidth{0pt}
\tablecaption{Flux Ratios of the Various Components of IRAS 17150$-$3224\label{fr150}}
\tablehead{
\colhead{Flux Ratio}&\colhead{Lobes}&\colhead{Arcs}&\colhead{Halo}&\multicolumn{2}{c}{Caps}&\colhead{Central Star}&\multicolumn{2}{c}{Torus} \\
\cline{5-6} \cline{8-9}
& & & &\colhead{NW}&\colhead{SE}& & \colhead{NE}&\colhead{SW}
}
\startdata
F160W/F606W&0.3$\sim$0.6&0.5&2.0&1.3&3.2&$>$65&24&33 \\
F222M/F160W&0.3 &\nodata& \nodata& 0.3&0.5&1.2& 1.2&1.3 \\
\enddata
\end{deluxetable}


\begin{deluxetable}{ccccc}
\tablewidth{0pt}
\tablecaption{Flux Ratios of the Various Components of IRAS 17441$-$2411\label{fr441}}
\tablehead{\colhead{Flux ratio}&\colhead{N Lobe}&\colhead{S Lobe}&\colhead{Halo}&\colhead{Central Star}
}
\startdata
F160W/F606W&1.8&2.3&1.8&$>$60\\
F222M/F160W&0.36&0.42&0.42&2.0 \\
\enddata
\end{deluxetable}


\begin{deluxetable}{cccccc}
\tablewidth{0pt}
\tablecaption{Maximum Observed Percentage Polarization Values\label{pmax}}
\tablehead{
\colhead{ }& \colhead{IRAS} & \colhead{IRAS} & \colhead{IRAS} & \colhead{IRAS} & \colhead { Egg Nebula} \\
\colhead{ } & \colhead{17150$-$3224} & \colhead{17441$-$2411} & \colhead{16594$-$4656} & \colhead{17245$-$3951} &
\colhead { }
}
\startdata
$p_{max}$(obs)  & 76 \%  & 60 \% & 60 \% & 42 \%  & 96 \% \\
Location   & fainter lobe & fainter lobe & fainter lobe & brighter lobe & fainter lobe \\
Size (\arcsec) & 9.3$\times$3.4 & 5.4$\times$2.7 &  8.4$\times$4.7 & 3.0$\times$2.0  & 12$\times$4 \\
\enddata
\end{deluxetable}

\clearpage

\begin{figure}
\figurenum{1}
\caption{Near-infrared images of IRAS 17150$-$3224: (a) F160W, (b) F222M,
        (c) flux-ratio map of F160W/F606W, and
        (d) flux-ratio map of F222M/F160W.
        The first three are displayed with a logarithmic scale and the fourth with a linear scale.
        In the flux-ratio maps, the redder regions are darker.
        The intensity range in (a) is 10$^{-13}$ to 10$^{-9}$ ergs cm$^{-2}$ s$^{-1}$ \um$^{-1}$ arcsec$^{-2}$
        and in (b) is 10$^{-14}$ to 10$^{-10}$ ergs cm$^{-2}$ s$^{-1}$ \um$^{-1}$ arcsec$^{-2}$.
        An inset in the F160W image uses intensity contours to show the region around the central star.}

\end{figure}

\begin{figure}
\figurenum{2} \caption{Radial intensity profiles of the arcs in
both the WFPC2 F606W and NIC2 F160W images of IRAS 17150$-$3224.
The first six arcs (A$-$F) seen in the visible WFPC2 image are
also clearly seen in the near-infrared NIC2 image. The arc F (peak
value) has a 3 sigma-detection in the de-convloved H image.} \label{150arc}
\end{figure}

\begin{figure}
\figurenum{3}
\caption{
Intensity profiles along three cuts in the F222M/F160W color map of IRAS 17150$-$3224.
Left: enlarged image of the central region with the three cuts indicated.
Right: intensity profiles along the three cuts. The central star is at zero radius: positive radii
represent the right side of the torus (SW) and negative radii represent the left side of the
torus (NE).  The position angles (P.A.) are measured with respect to
the horizontal direction. }
\label{150disk} \epsscale{1.1}
\end{figure}

\begin{figure}
\figurenum{4}
\caption{Near-infrared images of IRAS 17441$-$2411: (a) F160W, (b) F222M,
        (c) flux-ratio map of F160W/F606W, and (d) flux-ratio map of
        F222M/F160W.   All are displayed on a logarithmic scale.
        In the flux-ratio maps, the redder regions are darker.
        The intensity range in (a) is 10$^{-13}$ to 3$\times$10$^{-9}$ ergs cm$^{-2}$ s$^{-1}$ \um$^{-1}$ arcsec$^{-2}$
        and in (b) is 3$\times$10$^{-13}$ to 3$\times$10$^{-10}$ ergs cm$^{-2}$-s$^{-1}$ \um$^{-1}$ arcsec$^{-2}$.   An enlargement of the equatorial region
        of the F160W/F606W flux-ratio map with intensity contours is displayed as an inset.}
\label{441hk}
\end{figure}

\begin{figure}
\figurenum{5}
\caption{Near-infrared images of IRAS 17245$-$3951: (a) F160W, (b) F222M, and
        (c) flux-ratio map of F160W/F606W, all on a logarithmic scale.
        The redder regions are darker on the flux-ratio map.
         The intensity range in (a) is 10$^{-13}$ to 3$\times$10$^{-9}$ ergs cm$^{-2}$ s$^{-1}$ \um$^{-1}$ arcsec$^{-2}$
        and in (b) is 10$^{-13}$ to 3$\times$10$^{-10}$ ergs cm$^{-2}$ s$^{-1}$ \um$^{-1}$ arcsec$^{-2}$.}
\label{245hk}
\end{figure}

\begin{figure}
\figurenum{6}
\caption{Images of IRAS 16594$-$4656: (a) F160W, (b) F222M, (c) F606W,  and
(d) flux-ratio map of F160W/F606W, all on a logarithmic scale.
The F606W image is included to show the faint nebulosity seen in the visible image,
and it has the same orientation and scale as the F160W and F222N images.
The intensity range in (a) is 10$^{-14}$ to 3$\times$10$^{-8}$ ergs cm$^{-2}$ s$^{-1}$ \um$^{-1}$ arcsec$^{-2}$
 and in (b) is 3$\times$10$^{-14}$ to 10$^{-9}$ ergs cm$^{-2}$ s$^{-1}$ \um$^{-1}$ arcsec$^{-2}$.}
\label{594hk}
\end{figure}

\begin{figure}
\figurenum{7} \caption{The 2 \um polarimetric results of the PPNs.
The total intensity of the light through the three polarizers
(Stokes I) is plotted by contours and the polarimetric vectors are
plotted as lines whose lengths indicate the percentage
polarization. The contour levels range from 98$\%$ to 0.1$\%$ of
the peak values and are in equal intervals on a logarithmic scale.
A scale for the percentage polarization is given for each object.
The relative scales on the x and y axes are in arcsecs.}
\label{pol}
\end{figure}

\begin{figure}
\figurenum{8}
\caption{Polarimetric results for IRAS 17150$-$3224, displayed in different ranges of the
        percentage polarization: (a)  $70 \% > p > 60 \%$, (b) $60 \% > p > 50 \%$,
        (c) $50 \% > p > 40 \%$, (d) $40 \% > p > 30 \%$, (e) $30 \% > p > 20 \%$,
        (f) $20 \% > p > 10 \%$, (g) $p < 10 \%$.  These show a striking point symmetry.
        The intensity contours and x and y axis scales are as in Fig. 7.}
\label{pflowa}
\end{figure}

\begin{figure}
\figurenum{9}
\caption{Polarimetric results for IRAS 17441$-$2411, displayed in different ranges of the
        percentage polarization: (a) $p > 50 \%$, (b) $50 \% > p > 40 \%$,
        (c) $40 \% > p > 30 \%$, (d) $30 \% > p > 20 \%$, (e) $20 \% > p > 10 \%$,
        (f) $p < 10 \%$.  A general point symmetry is seen.
        The intensity contours and x and y axis scales are as in Fig. 7.}
\label{pflowb}
\end{figure}

\begin{figure}
\figurenum{10}
\caption{Polarimetric results for IRAS 17245$-$3951, displayed in different ranges of the
        percentage polarization: (a) $p > 30 \%$, (b) $30 \% > p > 20 \%$,
        (c) $20 \% > p > 10 \%$, (d) $p < 10 \%$.
        The intensity contours and x and y axis scales are as in Fig. 7.}
\label{pflowc} \epsscale{0.7}
\epsscale{1.0}
\end{figure}

\begin{figure}
\figurenum{11}
\caption{Polarimetric results for IRAS 16594$-$4656, displayed in different ranges of the
        percentage polarization: (a) $ 60 \% > p > 40 \%$, (b) $40 \% > p > 30 \%$,
        (c) $30 \% > p > 20 \%$, (d) $20 \% > p > 10 \%$, (e) $p < 10 \%$.
        The intensity contours and x and y axis scales are as in Fig. 7.}
\label{pflowd}
\end{figure}

\begin{figure}
\figurenum{12} \caption{Percentage polarization maps for  IRAS
17150$-$3224 and IRAS 17441$-$2411.}
\label{p_percent}
\end{figure}

\end{document}